\documentclass{article}

\usepackage{arxiv}

\usepackage[utf8]{inputenc} 
\usepackage[T1]{fontenc}    
\usepackage[numbers,sort&compress]{natbib}
\usepackage{amsmath}
\usepackage{graphicx}
\usepackage{hyperref}       
\usepackage{url}            
\usepackage{booktabs}       
\usepackage{amsfonts}       
\usepackage{nicefrac}       
\usepackage{microtype}      
\usepackage{lipsum}

\title{Homogeneous ice nucleation in an ab initio machine learning model of water}

\author{
    Pablo M. Piaggi \\
  Department of Chemistry, Princeton University, Princeton, NJ 08544, USA \\
  \texttt{ppiaggi@princeton.edu}
  \And
  Jack Weis \\
  Department of Chemical and Biological Engineering, Princeton University, Princeton, NJ 08544, USA \\  
  \And
  Athanassios Z. Panagiotopoulos \\
  Department of Chemical and Biological Engineering, Princeton University, Princeton, NJ 08544, USA \\
  \And
  Pablo G. Debenedetti \\
  Department of Chemical and Biological Engineering, Princeton University, Princeton, NJ 08544, USA \\
  \And
  Roberto Car \\
  Department of Chemistry, Princeton University, Princeton, NJ 08544, USA \\
  Department of Physics, Princeton University, Princeton, NJ 08544, USA\\
}

\begin{document}

\maketitle

\large

\begin{abstract}
Molecular simulations have provided valuable insight into the microscopic mechanisms underlying  homogeneous ice nucleation.
While empirical models have been used extensively to study this phenomenon, simulations based on first-principles calculations have so far proven prohibitively expensive.
Here, we circumvent this difficulty by using an efficient machine learning model trained on density-functional theory (DFT) energies and forces.
We compute nucleation rates at atmospheric pressure, over a broad range of supercoolings, using the seeding technique and systems of up to hundreds of thousands of atoms simulated with ab initio accuracy.
The key quantity provided by the seeding technique is the size of the critical cluster (i.e., a size such that the cluster has equal probabilities of growing or melting at the given supersaturation) which is used together with the equations of classical nucleation theory to compute nucleation rates. 
We find that nucleation rates for our model at moderate supercoolings are in good agreement with experimental measurements within the error of our calculation.
We also study the impact of properties such as the thermodynamic driving force, interfacial free energy, and stacking disorder on the calculated rates.
\end{abstract}

\newpage
Ice crystallization from supercooled liquid water is one of the most emblematic phase transformations to be found in nature.
It is of key importance in the regulation of our planet's climate\cite{Vergara18} and in many applications, such as artificial cloud seeding, cryopreservation, and food processing.
Molecular simulations have proven an invaluable tool to obtain insight into molecular-level details of this process and to make predictions at conditions not readily accessible to experiments.
For instance, Lupi et al.\ \cite{Lupi17} considered the effect of stacking disorder (i.e. the presence of alternate layers of hexagonal and cubic ice) on the nucleation rates and Sanz et al.\ \cite{Sanz13} used  systems of more than 100\,000 molecules in order to compute nucleation rates at low supercoolings.

However, simulations of ice nucleation carried out so far have employed relatively simple empirical models, such as the coarse-grained monoatomic model of water mW\cite{Molinero09} or the four-site rigid TIP4P water models\cite{Abascal05}.
A different route to study this phenomenon is using ab initio molecular dynamics\cite{Car85}.
In this technique, the forces acting on the atomic nuclei are derived from electronic structure calculations.
At variance with empirical models, the ab initio potential energy surface does not rely on empirical information, captures complex bonding behavior between atoms, and describes the formation and breaking of chemical bonds.
The solution of the many-body electronic Schr{\"o}dinger equation is, in general, not tractable and a widely-used approximation in this context is Kohn-Sham density-functional theory\cite{Kohn65} (DFT).
The application of ab initio molecular dynamics has, however, been limited for several decades to the simulation of relatively small systems ($\sim$ 1000 atoms) and short times ($\sim$ 100 ps) due to its high computational cost.
This limitation has precluded the study of ice nucleation from first principles.

A solution to this conundrum has been the use of machine learning algorithms that are able to learn the energies and forces derived from DFT data\cite{Behler07}.
The machine learning interatomic models constructed in this fashion reproduce the ab initio potential energy surface with high fidelity, are several orders of magnitude faster than DFT, and also show linear scaling with the number of nuclei.
Such models have recently been applied to the study of crystal nucleation in silicon\cite{Bonati18} and gallium\cite{Niu20}.
Previous simulations using first-principles models, however, explored only relatively large supercoolings for which systems of a few thousand atoms are able to contain the required crystalline cluster.

Here, we compute ice nucleation rates using an ab initio machine learning model of water.
We employ the seeding technique\cite{Sanz13} and systems of up to 300\,000 atoms in order to obtain nucleation rates in a broad range of supercoolings.
Our results allow us to compare predictions from a model derived from first principles with direct experimental measurements of nucleation rates.
Although we only simulate explicitly clusters of hexagonal ice, we take into account the effect of stacking disorder using a model for the chemical potential of ice with stacking disorder.

During homogeneous ice nucleation an ice cluster is formed within bulk liquid water.
Typically, this phenomenon takes place below the melting temperature and thus there is a driving force for the formation of ice.
However, the formation of an ice cluster in the liquid creates a liquid-solid interface with an associated energetic penalty.
The competition between a favourable bulk term and an unfavourable surface term leads to a free energy barrier that the system must surmount in order to proceed with the transformation.
The existence of a free energy barrier makes nucleation a rare event and severely hinders the ability to study the phenomenon directly using molecular simulations.
Although there have been attempts to study ice nucleation using straightforward molecular simulations\cite{Matsumoto02}, in general the problem must be tackled using more sophisticated techniques.

A possible route to study ice nucleation on the computer are rare event techniques, such as path sampling\cite{Bolhuis02,Lupi17}, forward flux sampling\cite{Li11,Haji15}, or metadynamics\cite{Laio02,Niu19}.
These approaches can provide valuable insights into the nucleation mechanism, albeit at a high computational cost.
A simpler alternative is the seeding technique\cite{Sanz13} that is based on performing a series of relatively short simulations at different temperatures starting from a configuration that contains an ice cluster embedded in liquid water.
The aim of these simulations is to find the temperature $T^*$ for which the chosen cluster is critical, that is to say, at $T^*$ the cluster has equal probabilities of growing and thawing.
This information is then used in combination with the equations of classical nucleation theory (CNT)\cite{KalikmanovBook} to calculate the nucleation rate.
This approach has several potential pitfalls that can affect the calculated rates, such as the appropriate choice of an order parameter to calculate the cluster size and the applicability of CNT to the nucleation process been studied.
These limitations have been carefully considered in the literature\cite{Zimmermann18} and the seeding technique has been shown to provide nucleation rates in good agreement with other methods\cite{Espinosa14,Lamas21}.

Another crucial ingredient in the simulation of ice nucleation is an accurate description of the interatomic interactions.
Here, we derive the forces between nuclei from first principles calculations.
In particular, we use density-functional theory (DFT) adopting the Strongly Constrained and Appropriately Normed\cite{Sun15} (SCAN) exchange and correlation functional.
SCAN is arguably one of the best semilocal functionals available and many properties of ice and water have been studied using this functional, e.g., in Refs.\ \citenum{Chen17} and \citenum{Piaggi21}.
Driving the dynamics directly using DFT forces would be unduly costly and instead we use a machine learning model trained on DFT data.
The model is based on deep neural networks and was constructed using the deep potential methodology developed by Zhang et al.~\cite{Zhang18}.
Below, we refer to this model as SCAN-ML (i.e., SCAN-trained, machine learning-based model).
The SCAN-ML model was carefully trained to reproduce data over a vast region of the phase diagram of water\cite{Zhang21}.
SCAN-ML has been used to provide evidence of the existence of a liquid-liquid transition at deeply supercooled conditions\cite{Gartner20} and to study the ice I$_{\mathrm{h}}$-ice XI transition\cite{Piaggi21b}.
The thermodynamic properties of this model relevant to ice nucleation were thoroughly characterized in ref.~\citenum{Piaggi21}.
The model has a melting temperature of 312 K, around 40 K larger than the experimental value.
The density change upon melting is 6\% in the model, somewhat smaller than the 9\% found in experiments.
Another important property is the relative stability between ice I$_{\mathrm{h}}$ and ice I$_{\mathrm{c}}$ that are the two competing polymorphs during ice nucleation at ambient pressure.
The SCAN-ML model correctly predicts that ice I$_{\mathrm{h}}$ is more stable than ice I$_{\mathrm{c}}$ in agreement with experiments.
Ref.~\citenum{Piaggi21} also analyzed the ability of the SCAN-ML model to reproduce SCAN energies in configurations that contain atomic environments compatible with both liquid water and ice, and found that the model is a faithful representation of SCAN with deviations of less than 1.3 meV per H$_2$O molecule.
We provide in Table \ref{tab:Table1} a summary of the properties of the SCAN-ML model and we compare them with experimental data and results using the empirical water models TIP4P/Ice and mW.

Before describing the results of our simulations, we briefly discuss the advantages of SCAN-ML over empirical models.
SCAN-ML is an all-atom fully-flexible model at variance with empirical potentials such as mW, which is a coarse-grained model, and TIP4P/Ice, which is an all-atom rigid model.
Since SCAN-ML reproduces the DFT potential energy surface, the flexibility of the OH bonds depends on the environments while in flexible empirical models, such as TIP4P/2005f\cite{Gonzalez11}, the flexibility of the bonds is modeled using simple functional forms and a few parameters that do not depend on the environment.
Another property that depends on the environment is the dipole moment of the water molecule.
For instance, the dipole moment is different in liquid water and ice\cite{Chen17}, but can also exhibit more subtle changes with the environment\cite{Zhang20dielectric,Sommers20}.
SCAN-ML is polarizable and able to capture the effects connected to changes in dipole moment\cite{Sommers20,Piaggi21b}. 
SCAN-ML is also fully reactive and can describe the proton transfer process in water.
This model captures many body interactions beyond 2- and 3-body, while mW is limited to 3-body interactions and TIP4P/Ice is based only on 2-body interactions.
SCAN-ML and TIP4P/Ice can both describe an important feature of ice I$_{\mathrm{h}}$, namely proton disorder, which is absent in the coarse-grained mW due to the lack of protons.

Our simulations based on SCAN-ML also have several limitations.
While the electronic degrees of freedom are treated quantum mechanically, the dynamics of the nuclei are based on the equations of motion of classical mechanics.
Therefore, we ignore nuclear quantum effects that could be modeled using path integral molecular dynamics.
Another disadvantage is that SCAN-ML is around 1 to 2 orders of magnitude more computationally expensive than empirical models. 
Also, the properties of SCAN-ML differ somewhat from experimental properties and this shows the limitations of the SCAN functional in the description of water and ice.
Lastly, the model is short-ranged, with an interaction cutoff of 6 \AA.
It thus can not capture the long range electrostatic interactions (present, for instance, in TIP4P models) nor van der Waals forces beyond this range.
Long-range electrostatic interactions could be modeled using the recently introduced deep potential long range (DPLR) scheme\cite{Zhang22}.

\begin{table}[b!]
    \centering
    \caption{Melting temperature ($T_m$), densities of ice I$_{\mathrm{h}}$ and liquid water at coexistence ($\rho_{\mathrm{ice}}$ and $\rho_l$), and enthalpy of fusion ($\Delta H_f$) of SCAN-ML, experimental water, and the empirical models TIP4P/Ice and mW.}
    \begin{tabular}{ccccc}
          &  $T_m$ (K) & $\rho_{\mathrm{ice}}$ (g/cm$^3$
) & $\rho_l$ (g/cm$^3$
) & $\Delta H_f$ (kJ/mol) \\
         \hline
         SCAN-ML \cite{Piaggi21} & 312(1) &  0.949(1) &  1.002(3) & 7.6(1) \\
         Experiment  & 273.15 & 0.917 & 0.999 & 6.01 \\
         TIP4P/Ice \cite{Abascal05}  & 270 & 0.906 & 0.985 & 5.40 \\
         mW \cite{Qiu18,Prestipino18} & 273 & 0.978 & 1.001 & 5.3 \\
    \end{tabular}
    \label{tab:Table1}
\end{table}

We now turn to discuss the results of the seeding simulations.
We studied ice I$_{\mathrm{h}}$ clusters of around 200, 700, and 4\,500 molecules embedded in liquid water and the corresponding total number of water molecules in the simulation boxes were around 4\,000, 12\,000, and 100\,000, respectively.
The choice of system size is discussed in detail in the Supplementary Information.
The initial, equilibrated configurations of such ice clusters are shown in Figure 1 a, b, and c.
We refer the reader to the Methods section for information about the equilibration procedure.
The clusters are nearly spherical, an observation that will be important when CNT is used to calculate several physical properties (see below).
Some faceting of the clusters can be observed and the hexagonal shape of the clusters is compatible with the six-fold symmetry of the basal plane of ice Ih.

\begin{figure}[htbp]
\includegraphics[width=\textwidth]{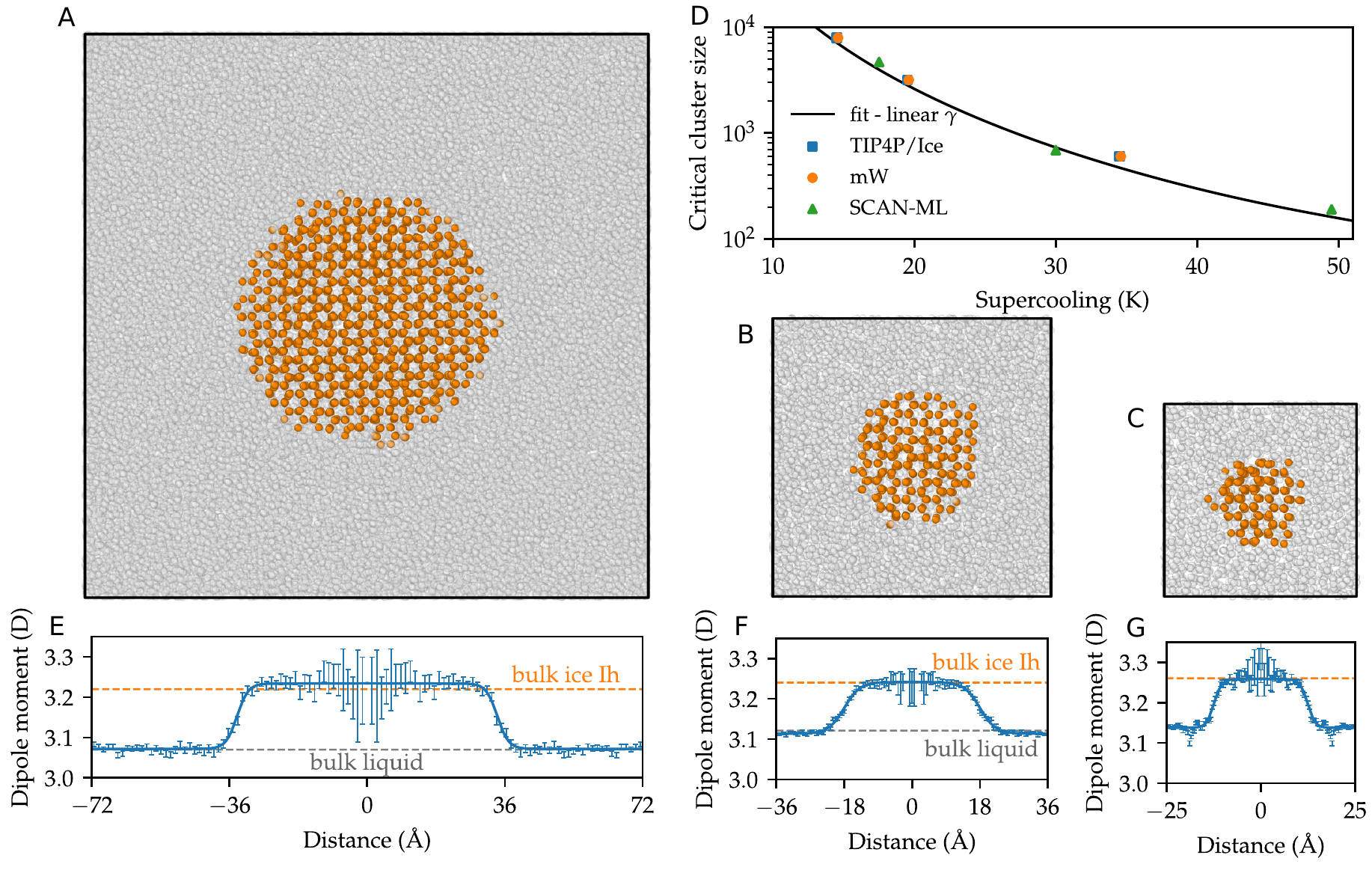}
\caption{\textbf{Ice I$_{\mathrm{h}}$ clusters employed in the seeding simulations.} a, b, and c show snapshots of the equilibrated cluster configurations in which only oxygen atoms are shown. Atoms with ice I$_{\mathrm{h}}$-like environments\cite{Larsen16,Stukowski09} are shown in orange and atoms with liquid-like environments are shown in gray. d shows the supercooling for which each of the clusters is critical in the SCAN-ML model. The curve labelled fit - linear $\gamma$ is based on the CNT formula $N^* = (32 \pi \gamma^3)/(3 \rho_{ice}^2 |\Delta\mu|^3)$ and uses the linear fit to $\gamma$ shown in Fig.~\ref{fig:Figure3}a. Results from ref.~\citenum{Espinosa16} for empirical models mW and TIP4P/Ice are also shown. e, f, and g show the quantum-mechanical average dipole moment of the water molecule as a function of the distance from the center of the ice clusters. Hyperbolic tangent functions were fit to the data and are shown in solid blue lines. Reference values for bulk ice I$_{\mathrm{h}}$ and liquid water were calculated at the equilibration temperatures (240 K, 275 K, and 290 K) and are shown with dashed orange and grey lines. The reference value for the liquid at 240 K is not provided due to the very long relaxation times at this temperature.}
\label{fig:Figure1}
\end{figure}

Molecular dynamics simulations were performed at different temperatures starting from the equilibrated configurations.
The change in cluster size as a function of time is shown in Fig. S1 for the three cluster sizes at different temperatures.
From these simulations we identified the temperatures $T^*$ at which these clusters have equal probabilities of growing and thawing.
In Figure 1d we show $T^*$ for the three cluster sizes studied here with the SCAN-ML model.
In order to determine the cluster size $N^*$ we must choose a local order parameter.
The results depend somewhat on this choice\cite{Zimmermann18} and here we employ a criterion similar to the one used by Espinosa et al.\ \cite{Espinosa14} in order to compare our results with the data reported therein (see Methods section for details about our criterion to identify ice-like molecules).
We also include in Figure 1d the comparison with the results of Espinosa et al.\ \cite{Espinosa16} for two widely used empirical models, namely, mW and TIP4P/Ice.
The results show that the critical cluster sizes are fairly independent of the model.
At the highest supercooling studied here, around 50 K, the dynamics of liquid water are very slow and thermal equilibration might not have been reached (see Supplementary Information for a detailed discussion on the relaxation times of liquid water in the SCAN-ML model).  

In order to illustrate the ability of SCAN-ML to capture subtle quantum-mechanical polarization effects, we calculated the average dipole moment of the water molecule as a function of the distance from the center of the ice I$_{\mathrm{h}}$ clusters.
The quantum-mechanical molecular dipole moment was computed according to the modern theory of polarization\cite{Resta92,King93}, adopting the formulation in terms of Wannier centers\cite{Resta94}.
The dependence of the Wannier centers on the coordinates of the atoms in the system was described by a deep neural network as described in ref.~\cite{Sommers20,Zhang20dielectric} (see Supplementary Information for further details of the calculation).
The results are shown in Fig.~\ref{fig:Figure1} e, f, and g.
The average dipole moment changes from around 3.25 D in the ice I$_{\mathrm{h}}$ cluster to around 3.1 D in the liquid water surrounding the cluster.
We also show in Fig.~\ref{fig:Figure1} e, f, and g the reference values for bulk ice I$_{\mathrm{h}}$ and liquid water (dashed lines) and the agreement with the dipoles in the cluster configurations is very good.
There is also good agreement between the bulk dipole moments calculated here and reference values obtained with SCAN DFT\cite{Chen17}.
The experimental dipole moment of liquid water at 298 K is  2.9 $\pm$ 0.6 D and is reproduced relatively well by SCAN\cite{Chen17}.
We note that the average dipole moment of the water molecule is a function of the temperature.
Since each cluster has been equilibrated at a different temperature, the reference bulk values differ in subplots e, f, and g of Fig.~\ref{fig:Figure1}.
Furthermore, in the configurations with the ice cluster embedded in liquid water, the average dipole moment transitions smoothly from the bulk ice I$_{\mathrm{h}}$ value to the bulk liquid value, and the water molecules at the interface have, on average, intermediate values of the dipole moment.

We now turn to assess the performance of the SCAN-ML model to describe ice nucleation.
We calculate nucleation rates by combining the information obtained from the seeding simulations with CNT.
The predictions of CNT rest on various assumptions\cite{KalikmanovBook}, for instance, CNT assumes that clusters are spherical and show bulk ice I$_{\mathrm{h}}$ properties.
Within CNT the nucleation rate (nuclei per unit time per unit volume) is,
\begin{equation}
    J = \rho_l Z f \exp(-\beta \: \Delta G ^* )
    \label{eq:rates}
\end{equation}
where $\rho_l$ is the density of the liquid, $Z$ is the Zeldovich factor which represents the probability of a critical cluster to cross the energy barrier, $f$ is the attachment rate, $\beta=1/(k_B T)$, $T$ is the temperature, and $k_B$ is the Boltzmann constant.
The nucleation free energy barrier, $\Delta G ^*$, can be calculated using the CNT formula,
\begin{equation}
\Delta G ^* = \frac{|\Delta\mu| \: N^*}{2},
\label{eq:barrier1}
\end{equation}
where $\Delta\mu$ is the difference in chemical potential between liquid water and ice I$_{\mathrm{h}}$, and $N^*$ is the number of water molecules in the critical cluster.
Eq.\ \eqref{eq:barrier1} provides a convenient way to calculate rates from $N^*$ and $T^*$ obtained in the seeding simulations.
$\rho_l$ and $\Delta\mu$ of the SCAN-ML model were calculated in ref.\ \cite{Piaggi21} and further details about the determination of $N^*$, $Z$, $f$, $\rho_l$ and $\Delta\mu$ can be found in the Methods section and in the Supplementary Information.

The nucleation rates thus calculated are shown in Figure 2 together with results of experiments\cite{Amaya18,Hagen81,Kramer99,Manka12,Murray10,Riechers13,Stan09}, and simulations using empirical models mW\cite{Espinosa16} and TIP4P/Ice\cite{Espinosa16,Haji15,Niu19}.
Most experiments are performed on micron-sized droplets and yield nucleation rates in the supercooling range 35-40 K\cite{Hagen81,Kramer99}.
However, experiments in the last 10 years have also used nano-sized droplets to reach much deeper supercoolings\cite{Amaya18,Manka12}.
We have not included in our plot the experimental results of Laksmono et al.\ \cite{Laksmono15} since there are discrepancies between their rates and most measurements.
Furthermore, it has been argued that more than one nucleus could have formed in those experiments\cite{Espinosa18}.
We have included in Figure 2 an horizontal line representing the experimental homogeneous nucleation limit, i.e. the rate at which a micron-sized droplet freezes in one second\cite{Espinosa14}.
We note that there are significant differences between the nucleation rates calculated from simulations using different methods.
For instance, in the case of TIP4P/Ice, estimates from forward flux sampling\cite{Haji15}, metadynamics\cite{Niu19}, and seeding\cite{Espinosa16} span about 10 orders of magnitude, which is, however, the typical error bar of the seeding technique.
Nucleation rates of the SCAN-ML model are in good agreement with experimental measurements within the uncertainty of our calculation.
Furthermore, the rates of SCAN-ML are intermediate between those of mW and TIP4P/Ice.
Therefore, the performance of SCAN-ML is similar to that of the best available semiempirical models.
As we shall see later, the inclusion of stacking disorder makes rates faster and reduces to some extent the discrepancy with experiment.
\begin{figure}[htbp]
\centering\includegraphics[width=0.5\textwidth]{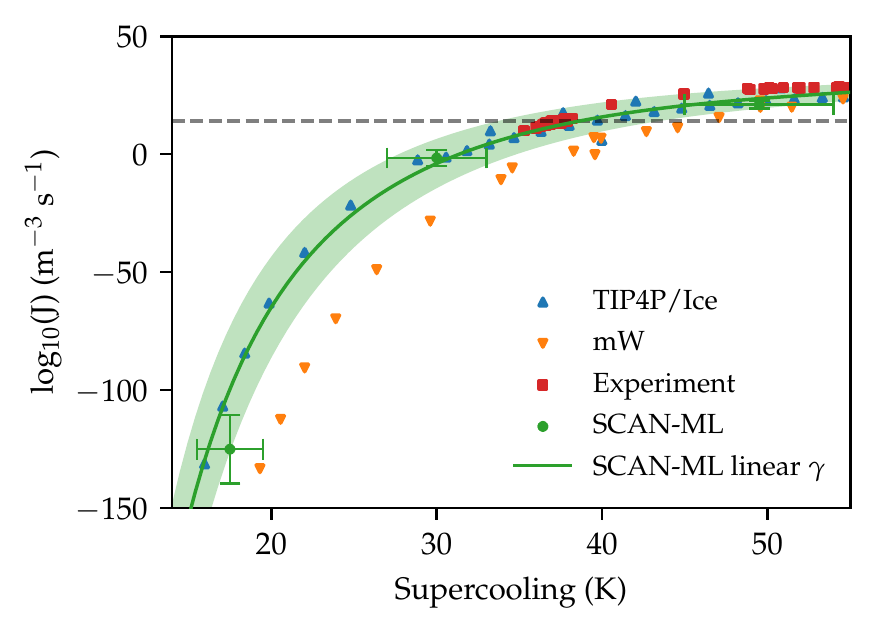}
\caption{\textbf{Ice nucleation rates as a function of supercooling.} Rates of the SCAN-ML model calculated in this work are compared with experimental data\cite{Amaya18,Hagen81,Kramer99,Manka12,Murray10,Riechers13,Stan09} and results from other works obtained using the models TIP4P/Ice\cite{Espinosa16,Haji15,Niu19} and mW\cite{Espinosa16,Haji14,Li11,Russo14}. We refer the reader to Figure S12 for further details about the computational techniques used to compute the rates of empirical models.
The solid green line labeled SCAN-ML linear $\gamma$ was obtained using the CNT Eq.~\eqref{eq:rates} and a linear fit to the interfacial free energy data presented in Fig.~\ref{fig:Figure3}.
The green shaded area is an estimate of the error in this calculation.
The experimental homogeneous nucleation limit\cite{Espinosa14} is shown as a horizontal gray dashed line and corresponds to log$_{10}$(J)(m$^{-3}$s$^{-1}$)=14. The calculation of error bars is described in the Supplementary Information.
}
\label{fig:Figure2}
\end{figure}

Another quantity that can be easily obtained from the seeding simulations is the interfacial free energy averaged over all orientations $\bar{\gamma}$.
For this purpose we employ the CNT expression,
\begin{equation}
    \bar{\gamma}=\left ( \frac{3 N^*}{32 \pi} \right )^{1/3} \rho_{\mathrm{ice}}^{2/3} |\Delta\mu|,
\end{equation}
where the symbols have the same meaning as in Eqs.\ \eqref{eq:rates} and \eqref{eq:barrier1}, and $\rho_{\mathrm{ice}}$ is the density of ice Ih.
The results of this calculation are shown in Figure 3a.
Data for the mW and TIP4P/Ice models obtained from seeding simulations\cite{Espinosa16} are also shown.
The dependence of $\bar{\gamma}$ on supercooling is a consequence of two different factors.
The first is that the interfacial free energy of a flat interface depends on the temperature  and the second is that each cluster has a different size and this will affect $\bar{\gamma}$ as shown by the Tolman equation\cite{KalikmanovBook}. 

\begin{figure}[htbp]
\centering\includegraphics[width=\textwidth]{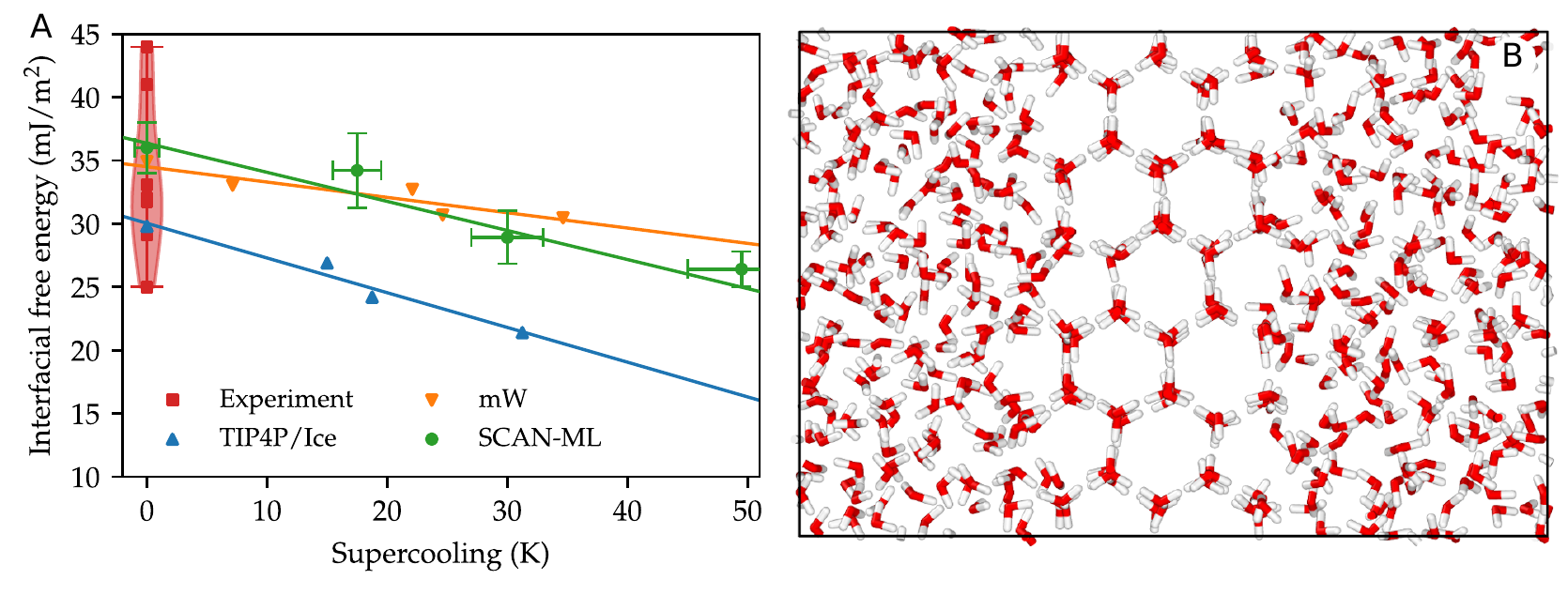}
\caption{\textbf{Liquid water-ice I$_{\mathrm{h}}$ interfacial free energy.} 
a) Interfacial free energy as a function of supercooling. Results for SCAN-ML at supercooling different from zero were obtained using data from the seeding simulations and assuming the validity of CNT. We have included data from Refs.\ \protect\citenum{Espinosa16,Espinosa16-surface} for the models mW and TIP4P/Ice, and experimental measurements at the melting temperature\protect\cite{Ickes15}. Linear fits to the results of the different models are shown as solid lines. The distribution of the experimental data is shown in red using a violin plot. The calculation of error bars is described in the Supplementary Information.
b) Planar interface between liquid water and the secondary prismatic plane of ice I$_{\mathrm{h}}$. This configuration was extracted from an advanced sampling simulation at 312 K driven by SCAN-ML during which the interface reversibly forms and melts.
}
\label{fig:Figure3}
\end{figure}

In order to validate the results obtained using seeding simulations, we also calculated the interfacial free energy $\gamma$ at coexistence for flat interfaces using advanced sampling simulations.
For this purpose, we computed $\gamma$ for the most relevant interfaces in ice I$_{\mathrm{h}}$, namely the prismatic ($1\bar{1}00$), secondary prismatic ($11\bar{2}0$), and basal ($0001$) planes.
The method to compute $\gamma$ for flat interfaces at coexistence is based on the reversible interconversion of the liquid and the respective liquid-ice I$_{\mathrm{h}}$ interface.
This is achieved by a suitably designed bias potential that increases the probability of observing the high free energy interfacial configuration.
A schematic of the interface sampled during the simulation of the secondary prismatic plane is shown in Figure 3b.
Further details of this approach and its validation can be found in the Methods section.
We also computed the interfacial free energy averaged over all orientations ($\bar{\gamma}$) as the mean of the three studied interfaces\cite{Espinosa14}.
The results of the free energy calculations are summarized in Table 1 and $\bar{\gamma}$ is shown in Figure 3a.
As seen in this figure, the agreement between $\bar{\gamma}$ obtained from advanced sampling calculations and seeding is very good.
For reference, we show in Table 2 results for the models mW and TIP4P/Ice as reported in ref.\ \citenum{Espinosa16-surface}.

\begin{table}[htbp]
    \centering
    \begin{tabular}{ccccc}
          & \multicolumn{4}{c}{Interfacial free energy (mJ/m$^2$)} \\
          & $\gamma_{(1\bar{1}00)}$ & $\gamma_{(11\bar{2}0)}$ & $\gamma_{(0001)}$ & $\bar{\gamma}$ \\
         \hline
         SCAN-ML & 36(2) & 34(2) & 37(2) & 36(2) \\
         Experiment \cite{Hardy77} & - & - & - & 29.1(8) \\
         Experiment (avg.) & - & - & - & $\sim$ 31.5 \\
         TIP4P/Ice \cite{Espinosa16-surface} & 31.6(8) & 30.7(8) & 27.2(8) & 29.8(8) \\
         mW \cite{Espinosa16-surface} & 35.1(8) & 35.2(8) & 34.5(8) & 34.9(8) \\
    \end{tabular}
    \caption{Interfacial free energy of ice I$_{\mathrm{h}}$ with liquid water at coexistence.  We report results for the prismatic ($1\bar{1}00$), secondary prismatic ($11\bar{2}0$) and basal ($0001$) planes. The interfacial free energy averaged over all orientations $\bar{\gamma}$ is also reported. We have included experimental results\cite{Ickes15,Hardy77} (see text for details) and calculations using the mW and TIP4P/Ice models\cite{Espinosa16-surface}.
    \label{tab:Table2}}
\end{table}

We have also included in Figure 3a and in Table 2 experimental results for $\bar{\gamma}$ at the melting temperature\cite{Ickes15}. 
There is no direct experimental measurement of $\bar{\gamma}$ at other temperatures and estimates based on CNT differ significantly\cite{Ickes15}.
For this reason we have not included them in our analysis.
The spread of the experimental results at the melting temperature is relatively large ($\sim$ 20 mJ/m$^2$) and has a mean value $\sim 31.5$ mJ/m$^2$ after removing outliers.
It has also been argued\cite{Espinosa14,Ickes15} that the experiments of Hardy\cite{Hardy77} based on the shape of the grain boundary groove provide the most reliable estimate, with a value of $29.1 \pm 0.8$ mJ/m$^2$.
$\bar{\gamma}$ for SCAN-ML is well within the region of uncertainty of the experimental measurements.
However, the interfacial free energy of SCAN-ML is higher than the average experimental estimate.
This behavior can be rationalized taking into account that SCAN-ML has a melting temperature and enthalpy of fusion higher than both the corresponding numbers for real water (experiments) and TIP4P/Ice.
Turnbull observed that there is a strong correlation between the interfacial free energy and the enthalpy of fusion\cite{Turnbull50}, and Laird has made a similar observation for the correlation between the interfacial free energy and the melting temperature\cite{Laird01}.
It is thus expected that the interfacial free energy of SCAN-ML should be higher than in the experiment and in TIP4P/Ice (melting temperature $\sim$270 K\cite{Abascal05}).
In Figure S4 we show that indeed the interfacial free energy correlates very well with the melting temperature in the TIP4P family and SCAN-ML.
An option to account for the different melting temperatures of the models is to compare $\bar{\gamma}$ using units of $k_B T$ for the energy.
A plot of $\bar{\gamma}$ in units of $k_B T / m^2$ vs supercooling is shown in Figure S11.
One could also estimate the value of $\bar{\gamma}$ in mJ/m$^2$ that SCAN-ML would have if its melting temperature were the experimental one.
An appropriate rescaling of $\bar{\gamma}$ is $\bar{\gamma}' = \bar{\gamma} \: T_m^{\mathrm{exp}} / T_m^{\mathrm{SCAN-ML}}$, where $T_m^{\mathrm{exp}}$ and $T_m^{\mathrm{SCAN-ML}}$ are the melting temperatures in the experiment and SCAN-ML, respectively.
In this way, we obtain a interfacial free energy for SCAN-ML of $\bar{\gamma}'=31.5$ mJ/m$^2$ at the melting temperature.

We now turn to analyze the thermodynamic properties of the models that affect nucleation rates by assuming the validity of CNT.
The nucleation barrier $\Delta G^*$ controls nucleation rates at low and moderate supercoolings since it is exponentiated in Eq.\ \eqref{eq:rates}.
The CNT expression for $\Delta G^*$ is,
\begin{equation}
\Delta G ^* = \frac{16 \pi \bar{\gamma}^3}{3 \rho_{\mathrm{ice} }^2|\Delta\mu|^2}.
\label{eq:barrier2}
\end{equation}
Therefore, the central physical quantities that govern nucleation rates at low and intermediate supercoolings are 1) the difference in chemical potential between liquid water and ice Ih ($\Delta\mu$), 2) the interfacial free energy of ice Ih with liquid water ($\bar{\gamma}$), and 3) the density of ice ($\rho_{\mathrm{ice}}$).
In the next paragraphs we analyze these quantities for SCAN-ML, TIP4P/Ice, and mW.

In Figure 4a we show the difference between $|\Delta\mu|$ in different models and in the experiment $|\Delta\mu^{\mathrm{exp}}|$ as a function of supercooling.
$|\Delta\mu^{\mathrm{exp}}|$ cannot be measured directly and its calculation from  experimentally measured heat capacities of liquid water and ice Ih\cite{Murphy05} is described in the Supplementary Information.
$|\Delta\mu|-|\Delta\mu^{\mathrm{exp}}|$ is reported in units of $k_B T$ in Figure 4a in order to compare models with different melting temperatures.
At 35 K of supercooling $|\Delta\mu|$ is underestimated by 9\% in SCAN-ML.
The performance of SCAN-ML in describing this property is somewhat better than that of TIP4P/Ice, which underestimates $\Delta\mu$ by 17\% at the same supercooling.
The mW model is the most accurate among the models considered here with $|\Delta\mu|$ at 35 K within 1\% of the experimental value.
However, mW changes from a underestimation of $|\Delta\mu|$ at low supercoolings to an overestimation at large supercoolings.
This is a consequence of a much weaker deviation of $\Delta\mu$ from a linear dependence with temperature than the other models (see Figure S8).
\begin{figure}[htbp]
\centering\includegraphics[width=0.5\textwidth]{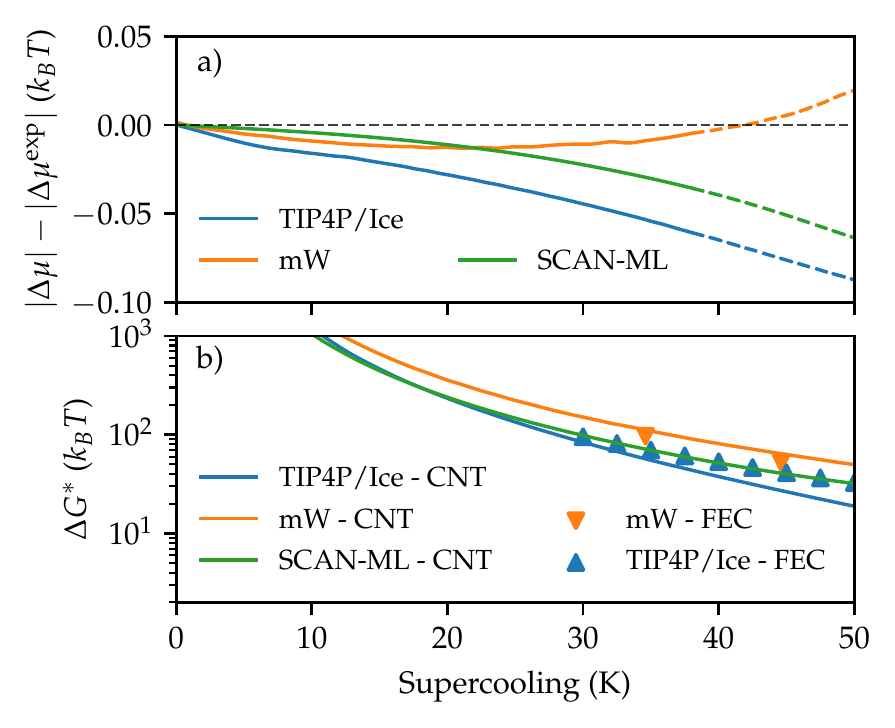}
\caption{\textbf{Analysis of the influence of supersaturation on nucleation barriers.} a) Difference between the driving force for nucleation in computer models $|\Delta\mu|$ and in the experiment $|\Delta\mu^{\mathrm{exp}}|$ as a function of supercooling. $|\Delta\mu|$ for the mW and TIP4P/Ice models was obtained from ref.~\citenum{Espinosa14}. The calculation of $|\Delta\mu^{\mathrm{exp}}|$ is based on experimental heat capacities\cite{Murphy05}. Above 38 K of supercooling results are shown as dashed lines to highlight the uncertainty in $|\Delta\mu^{\mathrm{exp}}|$ due to the lack of experimental measurements of the heat capacity of liquid water at these conditions. See SI for further details on the calculation of $|\Delta\mu|$. b) Nucleation barrier $\Delta G^*$ calculated using CNT (Eq.~\eqref{eq:barrier2} - see text for details). Results from umbrella sampling and metadynamics free energy calculations (FEC) reported in refs.~\citenum{Prestipino18,Cheng18,Niu19} are also shown.
}
\label{fig:Figure4}
\end{figure}

Results for the interfacial free energy were presented in Figure 3a.
There is limited experimental information to ascertain the deviation of the interfacial free energy with respect to experiments.
However, the values of $\bar{\gamma}$ for TIP4P/Ice and SCAN-ML (adjusted for the different melting temperature) are in relatively good agreement with most experimental results and most likely within a 5 \% error.
Instead, $\bar{\gamma}$ in the mW model is around 35 mJ/m$^2$ which is higher than the most reliable experimental estimates of $\bar{\gamma}$ and is most likely overestimated by around 10 \%.
It is also possible to characterize the temperature dependence of $\bar{\gamma}$ using the interfacial entropy, $S_{\gamma}=-\partial \bar{\gamma} / \partial T$, that can be estimated from the slope of $\bar{\gamma}$ with respect to temperature in Fig.~\ref{fig:Figure3}a.
We observe that mW has a lower slope $\partial \bar{\gamma} / \partial T$ than SCAN-ML and TIP4P/Ice, and that the latter two models have a similar slope.
This indicates that the interfacial entropy of the coarse-grained mW model is higher than in the TIP4P/Ice and SCAN-ML all-atom models that include protons explicitly.
Following ref.~\citenum{Qiu18}, the interfacial entropy can also be calculated using,
\begin{equation}
    S_{\gamma} \approx -\frac{\bar{\gamma}(T_m) \: \Delta C_p(T_m) }{\Delta H_f}
\end{equation}
where $\Delta H_f$ is the enthalpy of fusion and $\Delta C_p(T_m)$ is the difference in heat capacity between liquid water and ice I$_\mathrm{h}$ at the melting temperature $T_m$.
Using the values of $\Delta H_f$ and $\bar{\gamma}$ reported in Tables \ref{tab:Table1} and \ref{tab:Table2}, and $\Delta C_p(T_m)=49$ J/(mol K) \cite{Piaggi21} we obtain $S_{\gamma} \approx - 232$ $\mu$J m$^{-2}$ K$^{-1}$ for SCAN-ML.
This result is in good agreement with $S_{\gamma}$ calculated from experimental data (-215 $\mu$J m$^{-2}$ K$^{-1}$)\cite{Qiu18}.
A similar analysis for TIP4P/Ice gives $S_{\gamma} \approx - 226$ $\mu$J m$^{-2}$ K$^{-1}$ also in good agreement with the experiment.
The mW model has a $S_{\gamma}$ of -44 $\mu$J m$^{-2}$ K$^{-1}$ \cite{Qiu18} that is around a fifth of the experimental value.
These results are in agreement with the discussion above based on the slopes of the lines in Fig.~\ref{fig:Figure3}a.
From this discussion we deduce that an all-atom description seems essential to capture $\bar{\gamma}$ and its temperature dependence.

The density of ice I$_{\mathrm{h}}$ in the different models considered here is shown in Table \ref{tab:Table1} and in Fig.~S8 (data from refs.~\citenum{Prestipino18,Espinosa16,Rottger94}).
The density of SCAN-ML ice I$_{\mathrm{h}}$ is around 3 \% higher than in the experiment and according to \eqref{eq:barrier2} this would partially compensate the somewhat low $|\Delta\mu|$ in this model.
For TIP4P/Ice the density of ice I$_{\mathrm{h}}$ is around 1 \% lower than in the experiment and we expect it to have a negligible effect compared to other errors.
Finally, the mW model overestimates the density of ice I$_{\mathrm{h}}$ with respect to experiment by $\sim$ 7 \% and this might compensate in part for a large $\bar{\gamma}$.

We then computed the nucleation free energy barriers using Eq.\ \eqref{eq:barrier2}, the values for $\Delta\mu$ and $\rho_{\mathrm{ice}}$ reported in Figure S8, and the linear fits to $\bar{\gamma}$ shown in Figures 3a.
The results are shown in Figure 4b.
We have included barriers from refs.~\citenum{Prestipino18,Cheng18} and \citenum{Niu19} obtained using umbrella sampling and metadynamics free energy calculations.
For SCAN-ML we expect that the barrier should be overestimated since $|\Delta\mu|$ is underestimated.
This is compatible with the nucleation rates being somewhat slower than the experiment (see Figure 2).
In the TIP4P/Ice water model $|\Delta\mu|$ is underestimated more than in SCAN-ML and therefore we would expect an overestimation of the nucleation barrier and nucleation rates slower than in the experiment.
At variance with this prediction, the seeding nucleation rates\cite{Espinosa16} of TIP4P/Ice seem to agree relatively well with the experiments.
The rates calculated by Niu et al.\ \cite{Niu19} and Haji-Akbari and Debenedetti\cite{Haji15} for TIP4P/Ice, however, are slower than the experimental measurements.
In the case of the mW water model, $\Delta\mu$ is in very good agreement with the experiment.
For this reason, we surmise that the slow nucleation rates in this model can be traced back to an overestimation of $\bar{\gamma}$ not fully compensated by the overestimation of $\rho_{\mathrm{ice}}$.

Another important aspect of ice nucleation is stacking disorder.
There is significant experimental\cite{Amaya17} and computational\cite{Lupi17} evidence that nucleating ice clusters contain stacking faults, i.e. alternating layers of ice I$_{\mathrm{h}}$ and ice I$_{\mathrm{c}}$, and the solid polymorph that exhibits this feature is called ice I$_{\mathrm{sd}}$\cite{Malkin12}.
The prevalence of stacking faults in ice at equilibrium depends on two thermodynamic properties, namely, the difference in chemical potential between ice I$_{\mathrm{h}}$ and ice I$_{\mathrm{c}}$ $\Delta\mu_{I_{\mathrm{h}}\rightarrow I_{\mathrm{c}}}$, and the interfacial free energy between these two polymorphs $\gamma_{I_{\mathrm{h}}\rightarrow I_{\mathrm{c}}}$.
The experimental evidence on the value of $\Delta\mu_{I_{\mathrm{h}}\rightarrow I_{\mathrm{c}}}$ is limited due to the difficulty in obtaining pure ice I$_{\mathrm{c}}$, although very recently it has become possible to prepare samples with high structural purity\cite{delRosso20}.
The available experimental data puts $\Delta\mu_{I_{\mathrm{h}}\rightarrow I_{\mathrm{c}}}$ in the range from 0 to $\sim$ 200 J/mol (see ref.\ \citenum{Nachbar19} for a review).
An alternative point of view is provided by Lupi et al.~\cite{Lupi17} who argue that the experimental results reported in ref.~\citenum{Hondoh83} put an upper limit to $\Delta\mu_{I_{\mathrm{h}}\rightarrow I_{\mathrm{c}}}$ at $16.5 \pm 1.7$ J/mol.
On the computational side, the TIP4P/Ice and mW models have very small values of $\Delta\mu_{I_{\mathrm{h}}\rightarrow I_{\mathrm{c}}}$ of $\sim$ 0 \cite{Zaragoza15} and $\sim$ 5 J/mol\cite{Quigley14,Prestipino18}, respectively.
In ref.\ \citenum{Piaggi21} we have found a $\Delta\mu_{I_{\mathrm{h}}\rightarrow I_{\mathrm{c}}}$ for the SCAN-ML model of $65\pm37$ J/mol.
As we shall see, the precise value of $\Delta\mu_{I_{\mathrm{h}}\rightarrow I_{\mathrm{c}}}$ has an influence on rates, and further experimental and computational efforts are needed to shed light on its value.

We described the effect of stacking disorder using a model for the chemical potential of ice I$_{\mathrm{sd}}$ that rests on the following assumptions: 1) the entropy of mixing of ice I$_{\mathrm{c}}$ and ice I$_{\mathrm{h}}$ layers is ideal, 2) the interfacial free energy is negligible, and 3) stacking is only relevant in one direction, namely, the direction perpendicular to the basal plane of ice I$_{\mathrm{h}}$.
It can be shown that the first two assumptions give a lower bound for the chemical potential of ice I$_{\mathrm{sd}}$.
Since the effect of stacking disorder is more relevant when the chemical potential of ice I$_{\mathrm{sd}}$ is lower, then our model gives an upper bound for the possible effects of stacking disorder.
A more sophisticated 2D model has been used by Lupi et al.~\cite{Lupi17} and it was found that the simplified 1D model underestimates the entropic stabilization due to stacking disorder.
We also note that a similar model has been used by Pronk and Frenkel\cite{Pronk99}.
Further details can be found in the Materials and Methods section.
In Figure 5a we show the difference in chemical potential between ice I$_{\mathrm{sd}}$ and ice I$_{\mathrm{h}}$, $\Delta\mu_{I_{\mathrm{sd}}\rightarrow I_{\mathrm{h}}}$, as a function of supercooling as obtained from our model.
The model takes as input the difference in chemical potential between ice I$_{\mathrm{c}}$ and ice I$_{\mathrm{h}}$ $\Delta\mu_{I_{\mathrm{h}}\rightarrow I_{\mathrm{c}}}$.
We used two different values for this quantity, one compatible with the free energy of the mW model, $\Delta\mu_{I_{\mathrm{h}}\rightarrow I_{\mathrm{c}}}=5$ J/mol, and another one compatible with the SCAN-ML model, $\Delta\mu_{I_{\mathrm{h}}\rightarrow I_{\mathrm{c}}}=$ 65 J/mol.
In both cases $\Delta\mu_{I_{\mathrm{sd}}\rightarrow I_{\mathrm{h}}}$ becomes negligible as the supercooling goes to zero and the critical cluster size goes to infinity.
This reflects the fact that ice I$_{\mathrm{h}}$ is the most stable phase in the thermodynamic limit.
At larger supercoolings the finite size effects start to be important and ice I$_{\mathrm{sd}}$ becomes progressively more stable against ice I$_{\mathrm{h}}$.
The magnitude of the stabilization, around 50 J/mol, is  small compared with $|\Delta\mu|$ even at relatively large supercoolings.
\begin{figure}[htbp]
\centering\includegraphics[width=\textwidth]{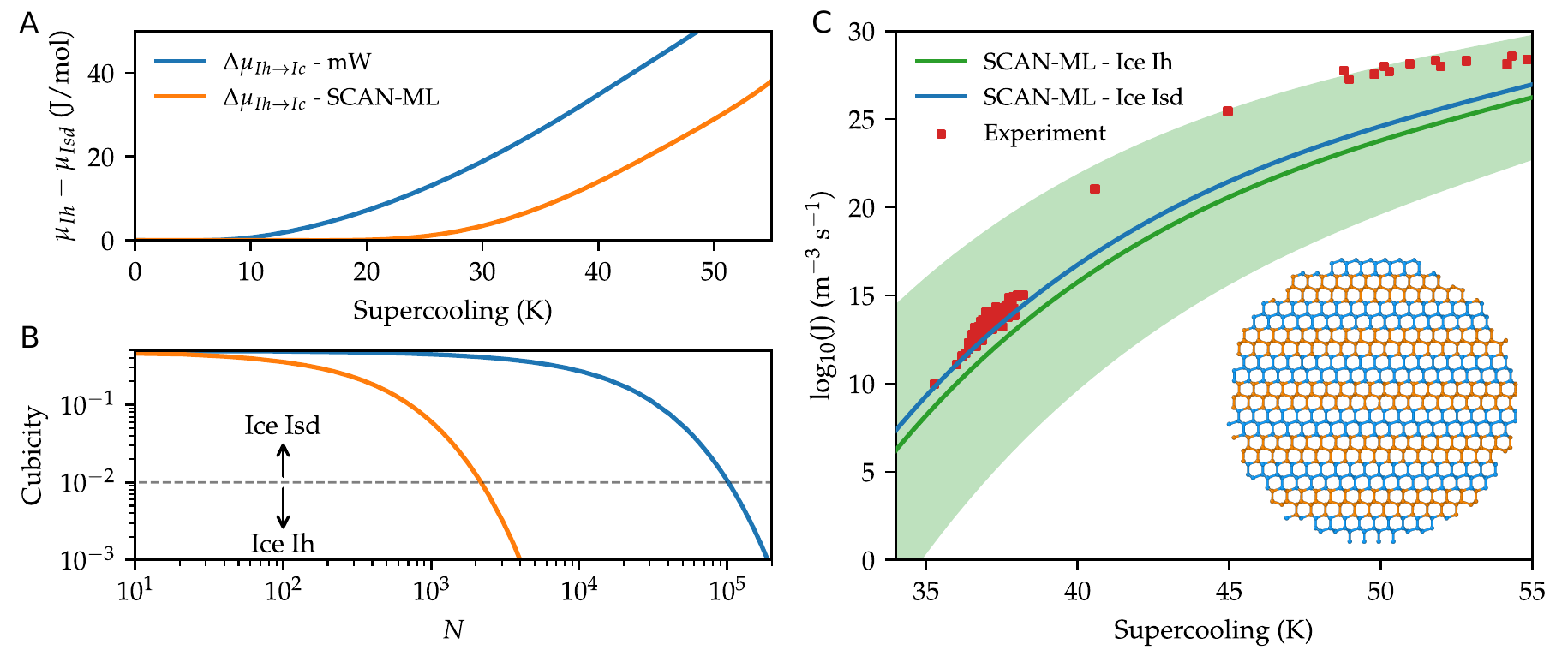}
\caption{\textbf{Influence of stacking disorder on the rates.} a) Difference in chemical potential between ice I$_{\mathrm{sd}}$ and ice I$_{\mathrm{h}}$,  $\Delta\mu_{I_{\mathrm{sd}}\rightarrow I_{\mathrm{h}}}=\mu_{I_{\mathrm{h}}}-\mu_{I_{\mathrm{sd}}}$, as a function of supercooling. b) Cubicity as a function of the number of molecules $N$ in a spherical ice cluster. The threshold to distinguish ice I$_{\mathrm{sd}}$ from ice I$_{\mathrm{h}}$ is shown with a dashed gray line and corresponds to a 1\% cubicity. c) Nucleation rates as a function of supercooling. Rates for ice I$_{\mathrm{h}}$ are compared with experimental data and the results for ice I$_{\mathrm{sd}}$ obtained with a model for stacking disorder. The green shaded region corresponds to the error in the rate of ice I$_{\mathrm{h}}$ and was calculated as described in the Supplementary Information. Two values are considered for $\mu_{I_{\mathrm{c}}}-\mu_{I_{\mathrm{h}}}$ in \eqref{eq:model}, one compatible with mW (5 J/mol) and another one compatible with SCAN-ML (65 J/mol).}
\label{fig:Figure5}
\end{figure}

It is also interesting to evaluate how $\Delta\mu_{I_{\mathrm{h}}\rightarrow I_{\mathrm{c}}}$ affects the cubicity of the nucleating clusters.
We evaluated the cubicity as a function of the size of the cluster and the results are reported in Figure 5b.
We found that for $\Delta\mu_{I_{\mathrm{h}}\rightarrow I_{\mathrm{c}}}=5$ J/mol the cubicity drops below 1\% for clusters of around 100\,000 molecules, in excellent agreement with findings of Lupi et al.\cite{Lupi17} for the mW water model.
Instead, for $\Delta\mu_{I_{\mathrm{h}}\rightarrow I_{\mathrm{c}}}=65$ J/mol the cubicity drops below 1\% at around 2\,000 molecules.
Therefore, the extent to which stacking disorder is relevant in small clusters depends largely on $\Delta\mu_{I_{\mathrm{h}}\rightarrow I_{\mathrm{c}}}$.

$\Delta\mu_{I_{\mathrm{sd}}\rightarrow I_{\mathrm{h}}}$ can be used within CNT to estimate the nucleation rates of ice I$_{\mathrm{sd}}$.
In Figure 5c we show the nucleation rates of ice I$_{\mathrm{sd}}$ calculated in this fashion.
For clarity we only show the results using a low value of $\Delta\mu_{I_{\mathrm{h}}\rightarrow I_{\mathrm{c}}}$, namely 5 J/mol, since this gives the greatest effect for the rates and is easier to visualize.
For this reason it should be considered an upper bound for the effect rather than the most reliable estimate.
In spite of the systematic choices we have made to obtain the maximum possible influence of stacking disorder on the rates, the effect of stacking disorder is around two or three orders of magnitude at deep supercoolings.
This relatively small change, however, improves somewhat the agreement of SCAN-ML with the experiment.
We also calculated nucleation rates using $\Delta\mu_{I_{\mathrm{h}}\rightarrow I_{\mathrm{c}}}=65$ J/mol and the results are shown in Figure S9.
This work shows that the latest advances in ab initio molecular dynamics allow studies of complex phenomena such as ice nucleation from first principles.
Our findings indicate that nucleation rates predicted based on SCAN DFT are in reasonably good agreement with experiment.
The rates are similar to those estimated with the TIP4P/Ice model and somewhat faster than the rates of the mW model.
The nucleation rate is a complex quantity that depends on many different properties of an atomistic model such as the density of liquid water and ice, the water-ice interfacial free energy, and the difference in chemical potential between water and ice.
We have performed a careful analysis of these properties and we have also compared them to the results of empirical models.
SCAN-ML gives a balanced description of these properties that results in good agreement between the calculated rates and the experimental measurements.

We have also highlighted limitations of the SCAN functional in describing some of the properties of liquid water and ice.
More accurate functional approximations and/or higher level quantum chemical data  are expected to improve the description of the properties of water.
It has also been shown that the MB-pol model\cite{Babin13} based on CCSD(T) calculations reproduces experimental properties with high accuracy and is thus an interesting model to study in the future.
Furthermore, in this work we have neglected nuclear quantum effects (NQE) that may have an important impact on some properties. For instance, the difference in chemical potential between liquid water and ice is influenced by heat capacities and the latter are affected significantly by NQE\cite{Vega10}.
In spite of their possible relevance, modeling NQE through path integral molecular dynamics is still computationally impractical for nucleation simulations that employ large system sizes such as the ones considered here. Understanding the impact of NQE on ice nucleation is an interesting direction for future work.
Finally, ab initio machine learning models of water can be extended to describe a substrate in order to simulate heterogeneous ice nucleation, a process of direct relevance to atmospheric science and climate modeling.
This would allow to include hitherto neglected phenomena, such as the effect of pH and the spontaneous hydroxylation of surfaces.
For these reasons, we foresee continued progress in the simulation of ice nucleation from first principles and the prediction of rates that are in progressively improved agreement with experiments as a result of an accurate description of the thermodynamic and kinetic properties of water and ice.

\section*{Methods}
\paragraph{Molecular dynamics} Simulations were performed using LAMMPS\cite{Plimpton95} patched with the DeePMD-kit\cite{Wang18}.
The temperature was kept constant with the stochastic velocity rescaling algorithm\cite{Bussi07} using a relaxation time of 0.1 ps.
A Parrinello-Rahman type barostat was used to maintain the pressure at 1 bar and a relaxation time of 1 ps was employed.
For the seeding simulations an isotropic barostat was used whereas for the advanced sampling simulations only the pressure component in the direction perpendicular to the interface was controlled.
The SCAN-ML model used with the DeePMD-kit was exactly the same as the one employed in ref.\ \citenum{Piaggi21}.
The performance of the implementation that we used was around 1 ns/day using an optimal number of GPUs for a given system size. The performance with the latest version of DeePMD-kit would have been faster at around 10 ns/day.

\paragraph{Seeding simulations} Configurations for the seeding simulations were constructed in the following way.
A simulations box with water molecules was prepared and then a spherical cavity was carved from its center.
This region was later filled with a seed of ice I$_{\mathrm{h}}$ with proton disorder created using GenIce\cite{Matsumoto18}.
This procedure was repeated for three different system sizes.
These systems contained 3\,934, 11\,872, and 99\,404 water molecules, respectively.
Afterwards the energy was minimized to a relative accuracy $10^{-6}$ and a 1 ns MD simulation for equilibration was performed at a temperature below the one for which the cluster is critical in order to avoid partial melting of the cluster.
The corresponding temperatures were 240, 275, and 290 K for the smallest, intermediate, and largest cluster, respectively.
 After equilibration, MD simulations were run at different temperatures to find $T^*$.
The simulations for the largest system were run on the Summit supercomputer using 600 Nvidia V100 graphical processing units (GPUs).
Each of the simulations would have required $\sim$5 years to be completed in a single GPU.
The intermediate and small system sizes used 100 and 24 GPUs per simulation, respectively.

The size of the clusters was determined using the local Steinhardt parameter $\bar{Q_6}$ proposed by Lechner and Dellago\cite{Lechner08}.
$\bar{Q_6}$ was calculated using the Freud\cite{Ramasubramani20} Python library v2.7.0.
The threshold value of $\bar{Q_6}$ that separates liquid and ice I$_{\mathrm{h}}$ environments was determined for each temperature using the criterion that an environment with the threshold value of $\bar{Q_6}$ has equal probabilities of being classified as liquid or ice I$_{\mathrm{h}}$.
Probability densities of $\bar{Q_6}$ for the liquid and ice I$_{\mathrm{h}}$ at different temperatures are shown in Fig.~S5a.
The chosen thresholds as a function of temperature are shown in Fig.~S5b and show a linear correlation.
We used a linear fit to these data to determine the threshold at any intermediate temperature.
All seeding simulations were analyzed using a threshold appropriate for the temperature of the simulation.
We note that the overlap of the liquid and ice I$_{\mathrm{h}}$ distributions increase sharply upon crossing the Widom line.
This can be expected given that below the Widom line liquid water resembles the low density liquid (LDL) water phase and ice I$_{\mathrm{h}}$ is known to be more similar to LDL than to the high density liquid (HDL). 
In Figure 1 the classification in ice-like and liquid-like environments was performed with the Polyhedral Template Matching algorithm\cite{Larsen16} as implemented in \textsc{Ovito}\cite{Stukowski09}.

The prefactor in the CNT expression for the nucleation rates (Eq.~\eqref{eq:barrier1}) requires the calculation of the Zeldovich factor $Z$ and the attachment rate $f$.
$Z$ was calculated using,
\begin{equation}
    Z = \sqrt {\frac{|\Delta\mu|}{6 \pi k_B T N^*}}
\end{equation}
and $f$ was calculated using\cite{Espinosa14},
\begin{equation}
    f = \frac{\langle \left (N(t)-N(0) \right )^2 \rangle}{2 t}
\end{equation}
where $N(t)$ is the cluster size at time $t$.
In practice, we computed $f$ from the slope of $\langle \left (N(t)-N(0) \right )^2 \rangle$ vs $2t$.
Results for $f$ are shown in Figure S7.

\paragraph{Advanced sampling simulations} The calculation of the ice I$_{\mathrm{h}}$-liquid water interfacial free energy was performed with LAMMPS augmented with the PLUMED enhanced sampling plugin\cite{Tribello14,Bonomi19}.
The initial configuration was made using GenIce and consisted in 288 water molecules in the ice I$_{\mathrm{h}}$ structure with proton disorder.
An equilibration of 1 ns at 312 K and 1 bar was then carried out, and the box dimensions were set to their average values during this run.
In order to obtain simulation boxes adequate for the simulation of the prismatic, secondary prismatic, and basal interfaces, the box was replicated along one of the three main axis and then the solid configuration was melted in a 1 ns run at 450 K while only the direction along which the box was replicated was barostated.

Next, we performed an advanced sampling simulation for each interface in which a bias potential was constructed using the On-the-fly Probability Enhanced Sampling (OPES) method\cite{Invernizzi20rethinking}.
This method is an evolution of the well-known metadynamics technique\cite{Laio02}.
The OPES bias potential was built as a function of a collective variable (CV) that counts the number of environments compatible with ice I$_{\mathrm{h}}$ in a region around an arbitrarily chosen atom (for instance, atom number 1).
The number of environments compatible with ice I$_{\mathrm{h}}$ was calculated using the environment similarity\cite{Piaggi19b} metric taking the four tetrahedral reference environments of ice I$_{\mathrm{h}}$ $\chi_i$ with $i=1,..,4$.
As a result of the introduction of the bias potential, during the biased simulations a slab of the ice I$_{\mathrm{h}}$ crystal is reversibly formed and melted.
The free energy difference between the liquid and the slab was calculated using,
\begin{equation}
\Delta G = - k_B T \log \left ( \frac{Z^{\ddag}}{Z^l}  \right )
\end{equation}
where $Z^{\ddag}$ and $Z^l$ are the partition functions of the slab and the liquid.
The interfacial free energy can then be calculated as,
\begin{equation}
\gamma = \frac{|\Delta G|}{2 A}
\end{equation}
with $A$ the cross section of the interface.
Further details are provided in the Supplementary Information.

This approach was validated by calculating the interfacial free energy of TIP4P/Ice that is known from literature\cite{Espinosa16-surface}.
The interfacial free energy of TIP4P/Ice averaged all interfaces was found to be 31(1) mJ/m$^2$ in good agreement with the estimate from literature 29.8(8) mJ/m$^2$. 

\paragraph{Model for stacking disorder} Stacking disorder was modeled using the following expression for the difference between the chemical potential of ice I$_{\mathrm{sd}}$ and I$_{\mathrm{h}}$,
\begin{equation}
    \mu_{I_{\mathrm{sd}}}(C,N)-\mu_{I_{\mathrm{h}}} = C (\mu_{I_{\mathrm{c}}}-\mu_{I_{\mathrm{h}}}) - \frac{1}{N} T S_{\mathrm{mix}} + \frac{1}{N} \sum_i \gamma_{sf} A_i
    \label{eq:model}
\end{equation}
where $C$ is the cubicity, $N$ is the number of molecules, the index $i$ runs through the ice I$_{\mathrm{h}}$-ice I$_{\mathrm{c}}$ interfaces, $\gamma_{sf}$ is the interfacial free energy of the stacking faults, and $A_i$ is the area of the $i$-th interface.
The first term in Eq.~\eqref{eq:model} is the bulk contribution of ice I$_{\mathrm{h}}$ and ice I$_{\mathrm{c}}$.
The second term is the contribution from the entropy of mixing of the stacked layers.
We assume that stacking is relevant only in one direction, i.e. the direction perpendicular to the basal plane of ice Ih.
The last term in Eq.~\eqref{eq:model} takes into account the penalty to form an ice I$_{\mathrm{h}}$-ice I$_{\mathrm{c}}$ interface.
The second and third terms go to zero as $N\to \infty$, reflecting that stacking disorder is only relevant for finite systems.

Since the effect of stacking disorder on the rates and chemical potentials is small, we do the following approximations.
First, we neglect the third term in Eq.\ \eqref{eq:model} that is always positive.
Second, we approximate the entropy of mixing with the ideal entropy of mixing,
\begin{equation}
    S_{\mathrm{mix}} \approx - N_l k_B (C \log(C)+ (1-C) \log(1-C))
\end{equation}
where $N_l$ is the number of stacked layers.
The ideal entropy of mixing is always larger than $S_{\mathrm{mix}}$.
These choices give a lower bound for $\mu_{I_{\mathrm{sd}}}(C,N)-\mu_{I_{\mathrm{h}}}$ and thus the greatest possible influence on the rates.
We make the additional assumption that the cluster of $N$ molecules is approximately spherical and calculate $N_l$ using the expression,
\begin{equation}
    N_l(N) = \frac{D}{d} = \left ( \frac{6 N}{\pi \rho_{ice}} \right )^{1/3} \frac{1}{d}
\end{equation}
where $D$ is the diameter of the cluster and $d$ is the distance between layers of the basal plane.
The cubicity and chemical potential in equilibrium are found by minimizing $\mu_{I_{\mathrm{sd}}}(C,N)$ with respect to $C$.
In order to obtain $\mu_{I_{\mathrm{sd}}}(C,N)$ as a function of temperature, we replace $N$ with the number of molecules $N^*_{I_{\mathrm{sd}}}(T)$ in a critical cluster with stacking disorder at a given temperature $T$.
$N^*_{I_{\mathrm{sd}}}(T)$ is not known but can be approximated by the number of molecules $N^*_{I_{\mathrm{h}}}(T)$ in a critical cluster of ice I$_{\mathrm{h}}$ that has been computed using seeding simulations.

\section*{Acknowledgments}

This work was conducted within the center Chemistry in
Solution and at Interfaces funded by the DoE under Award DE-SC0019394.
P.M.P was also supported by an Early Postdoc.Mobility
fellowship from the Swiss National Science Foundation.
This research used resources of the Oak Ridge Leadership Computing Facility at the Oak Ridge National Laboratory, which is supported by the Office of Science of the U.S. Department of Energy under Contract No. DE-AC05-00OR22725.
Simulations reported here were substantially performed using the Princeton Research Computing resources at Princeton University which is consortium of groups including the Princeton Institute for Computational Science and Engineering and the Princeton University Office of Information Technology’s
Research Computing department.

\section*{Data availability}
Input and output files of the simulations reported here are openly available on \url{https://doi.org/10.34770/xrd9-3d18} and on PLUMED-NEST (\url{https://www.plumed-nest.org/}), the public repository of the PLUMED consortium, as plumID:22.016.

\section*{Code availability}
LAMMPS, Plumed, and DeepMD are free and open source codes available at \href{https://
lammps.sandia.gov}{https://
lammps.sandia.gov}, \href{https://www.plumed.org}{https://www.plumed.org}, and \href{http://www.deepmd.org}{http://www.deepmd.org}, respectively.

\section*{Author contributions}
P.M.P, A.Z.P, P.G.D. and R.C. conceived the project; P.M.P and J.W. performed research; P.M.P, J.W., A.Z.P, P.G.D. and R.C. designed research, discussed results, and wrote the paper.

\end{document}